\documentclass[12pt]{article}
 
\begin{document}

\baselineskip=.2in
\lineskip=14pt plus 0.2pt minus 0.2pt

\newcommand{\be}{\begin{equation}}
\newcommand{\ee}{\end{equation}}
\newcommand{\bc}{\begin{center}}
\newcommand{\ec}{\end{center}}
\newcommand{\bea}{\begin{eqnarray}}
\newcommand{\eea}{\end{eqnarray}}
\newcommand{\da}{\dagger}
\newcommand{\dg}[1]{\mbox{${#1}^{\dagger}$}}
\newcommand{\hlf}{\mbox{$1\over2$}}
\newcommand{\lfrac}[2]{\mbox{${#1}\over{#2}$}}
\newcommand{\nsz}[1]{\mbox{\normalsize ${#1}$}}
\newcommand{\ep}{\epsilon(t)}
\newcommand{\epo}{\epsilon_o}
\newcommand{\eps}{\epsilon^*(t)}
\newcommand{\epso}{\epsilon_o^*}
\newcommand{\epd}{\dot{\epsilon}(t)}
\newcommand{\epdo}{\dot{\epsilon}_o}
\newcommand{\epsd}{\dot{\epsilon}^*(t)}
\newcommand{\epsdo}{\dot{\epsilon}_o^*}
\newcommand{\QR}{\langle Q \rangle}
\newcommand{\QRT}{\langle Q^2 \rangle}
\newcommand{\PR}{\langle P \rangle}
\newcommand{\PRT}{\langle P^2 \rangle}

 \begin{flushright}
quant-ph/0002050 \\
LA-UR-00-635 \\
\end{flushright} 

\textwidth=6.25in

\begin{quote}

\begin{flushleft}

\Large{\bf \textsf{Coherent states sometimes look like squeezed states,
and visa versa: The Paul trap}}\footnote{Our discussion of the physical 
distinction between coherent and squeezed states is most appropriate 
as a vehicle for us to pay homage to the late Dan Walls.  Since the 
time of their experimental discovery in the mid 1980s, his name has 
been closely associated with the physics of squeezed states.}
 \\
\vspace{0.25in}

\large

{\bf \textsf{Michael Martin Nieto}$\dagger$ 
 \textsf{and D. Rodney Truax}$\ddagger$}

\normalsize

\vspace{0.25in}

$\dagger$Theoretical Division (MS-B285), Los Alamos National Laboratory, 
University of California, Los Alamos, New Mexico 87545, U.S.A. \\
$\dagger$Universit\"at Ulm, Abteilung f\"ur Quantenphysik, 
Albert-Einstein-Allee \\ 11, D 89069 Ulm, Germany\\ 
$\ddagger$Department of Chemistry, University of Calgary, 
Calgary, Alberta T2N 1N4, Canada\\

\vspace{.1in}  
$\dagger$Email:  \texttt{mmn@lanl.gov} \\
$\ddagger$Email:  \texttt{truax@ucalgary.ca}

\vspace{.1in}  
\today

\vspace{0.3in}

{\bf  \textsf{Abstract.}} Using the Paul Trap as a model, we point out 
that the same wave functions 
can be variously coherent or squeezed states, depending upon the system they
are applied to.    

\end{flushleft}

\end{quote}



\section{\textsf{Introduction}}

Elsewhere \cite{oldI-II}-\cite{newII-III}, we have investigated 
time-dependent  Schr\"odinger 
equations that are quadratic in position, $q$, and momentum, 
$p=-i\partial_q$:
\bea
\left\{2i\partial_t \right.&-& \left[1+g_2(t)\right]p^2 
         + g_0(t)\lfrac{1}{2}(qp+pq)  \nonumber \\
  &+& \left.  g_1(t)p  -2h_2(t)q^2-2h_1(t)q
            -2h_0(t)\right\}\Psi(q,t) = 0. \label{segen}
\eea
We use dimensionless variables ($\hbar=m=1$) and the 
$g$'s and $h$'s are time-dependent functions. This study yielded methods 
for obtaining the symmetries of these systems.  It also provided explicit 
analytic relationships between various subclasses of these 
equations.  From that we could obtain implicit and sometimes explicit 
analytic solutions for the number, coherent, and squeezed states, and also
for the associated uncertainty relations.   

In this paper, we will apply the techniques so obtained to the Paul trap
system.  We will thereby be able to provide insight into what are the
coherent and squeezed states of the Paul trap, and into what uncertainty
relations are satisfied.  

In Section 2, we will set up the formalism for the particular subclass of 
Eq. (\ref{segen}) that contains the Paul trap as a special case.  That is: 
all $h_i(t) = g_i(t) = 0$ except $h_2(t) \equiv f(t)$.  We go on, in
Section 3, to describe the coherent states, squeezed states, and uncertainty
relations for this subclass.  The physical Paul trap is described 
in Section 4. 

In Section 5, we  discuss coherent states, squeezed states, and
uncertainty relations for the Paul trap. This is the main thrust of this 
paper.  Specifically, 
the reader will see how the same physical states can be viewed either as  
coherent states of the Paul trap or as squeezed states of the harmonic 
oscillator, and vice-versa.


\section{\textsf{Time-Dependent Quadratic Systems}}
\label{notation}

We now concentrate on the aforementioned 
particular subclass of Eq. (\ref{segen}), the time-dependent, 
harmonic-oscillator, Schr\"odinger equation. 
We have that\footnote{The 
Hamiltonian is $H=-\frac{1}{2}\partial_{qq}+f(t)q^2$.}
\be
\left\{\partial_{qq} + 2i\partial_t -2f(t)q^2\right\}\Psi(q,t)
=0.\label{SE}
\ee
Here the function $f(t)$ is a continuously differentiable and
integrable function of $t$.  From the general results
\cite{oldI-II}-\cite{newII-III} for Eq. (\ref{segen}),  we have that 
Eq. (\ref{SE}) admits a symmetry algebra that is a product of a 
Heisenberg-Weyl algebra, $w_1^c$, with a $su(1,1)$ algebra.  The 
subalgebras $w_1^c$ and $su(1,1)$ are the ones associated with coherent
states and  squeezing.   

The time-dependent lowering and raising (ladder) operators in $w_1^c$ 
are given by 
\be
A(t) = i\left[\xi(t)~p -\dot{\xi}(t)~q\right], ~~~~
A^\da(t) = -i\left[\xi^*(t)~p -\dot{\xi}^*(t)~q\right]. \label{ALADDER}
\ee
Here $q$ and $p = -i\partial_q$ are the ordinary position and momentum
operators
\be
q=\frac{a+a^{\da}}{\sqrt{2}},~~~~p=\frac{a-a^{\da}}{i\sqrt{2}}.
\label{aladder}
\ee
It is straight forward to demonstrate  \cite{oldI-II}
that the time-dependent functions, $\xi$ and $\xi^*$ 
in Eq. (\ref{ALADDER}) are two linearly independent, 
complex solutions of the second-order differential equation\footnote{
The `dot' over a function of $t$ indicates ordinary differentiation with 
respect to $t$.} 
\be
\ddot{\gamma}+2f(t)\gamma=0.\label{ode}
\ee
Therefore, the time-dependent functions $\xi$ and $\xi^*$ are determined 
by the particular Hamiltonian. 

The Wronskian of these solutions is 
\be
\xi(t)\dot{\xi}^*(t)-\dot{\xi}(t)\xi^*(t)=-i.\label{cw}
\ee
On account of Wronskian (\ref{cw}), the ladder operators $A(t)$ and 
$A^{\da}(t)$ satisfy the commutation relation 
\be
[A(t),A^{\da}(t)]=I,\label{com1}
\ee
where $I=1$ is the identity operator.


\section{\textsf{Coherent/squeezed states and  uncertainty relations}}
\subsection{\textsf{Coherent states}}

First, consider $|\alpha;t\rangle$,
the displacement-operator coherent states (DOCS).  
They are defined by  
\be
|\alpha;t\rangle = D_A(\alpha)|0;t\rangle = \exp\left[\alpha A^{\da}(t) - 
\alpha^* A(t)\right]|0;t\rangle.\label{docs}
\ee
In the above, $D_A(\alpha)$ is the displacement operator and $\alpha$ is a
complex   constant.
Similarly, the equivalent ladder-operator coherent states (LOCS) are defined
as 
\be
A(t)|\alpha;t\rangle = \alpha|\alpha;t\rangle.\label{locs} 
\ee

In Eq. (\ref{docs}), the extremal state $|0;t\rangle$ is a member of a 
set of number states, $\{|n;t\rangle,~n=0,1,2,\cdots\}$, 
which are eigenfunctions 
of the number operator $A^{\da}(t)A(t)$ {\cite{oldI-II,nt1}}:  
\be
A^{\da}(t)A(t)|n;t\rangle = n|n;t\rangle.\label{eig1}
\ee  
An important conceptual point for these time-dependent systems is that  
the number states, $|n;t\rangle$, are in general NOT eigenstates of the 
Hamiltonian.  That is,
\be 
H|n;t\rangle = i\partial_t|n;t\rangle \ne 
         {\rm (real~constant)}|n;t\rangle . 
\ee
In particular, the extremal state $|0;t\rangle$ is technically NOT the 
ground state.


\subsection{\textsf{Uncertainty relations and squeezed states}}

The operators, $A$ and $A^\da$, are linear functionals 
of $q$ and $p$.  With them, generalized position 
and momentum operators can be defined:
\be
Q = \frac{A + A^\da}{\sqrt{2}}, ~~~~ P = \frac{A -  A^\da}{i\sqrt{2}}.
\label{QP}
\ee
Their associated commutation relation is 
\be
[Q,P] = i{\cal O} = i[A,A^{\da}],
\ee
where ${\cal O}$ is an Hermitian operator.\footnote{ 
Because of Eq. (\ref{com1}), for the systems studied in this  paper 
${\cal O}=I$.  But for  now we continue with the general ${\cal O}$.}

Now consider the uncertainty product 
\bea
U(Q,P) = 
(\Delta Q)^2(\Delta P)^2&=& [\langle Q^2\rangle - \langle Q\rangle^2]
                          [\langle P^2\rangle - \langle P\rangle^2] \\
&=& [\langle (Q  - \langle Q\rangle)^2 \rangle]
            [\langle (P  - \langle P\rangle)^2 \rangle]. \label{upgen}  
\eea
Applying the Schwartz inequality to Eq. (\ref{upgen}) we have  
\bea
 (\Delta Q)^2(\Delta P)^2
  &\ge & \lfrac{1}{4}|i\langle {\cal O} \rangle 
          +  \langle \{ \hat{Q},\hat{P}\} \rangle|^2  \\   
&\ge & \lfrac{1}{4}\langle {\cal O} \rangle^2 
         + \lfrac{1}{4}\langle \{\hat{Q},\hat{P}\} \rangle^2  
\label{sur}  \\
&\ge & \lfrac{1}{4}\langle {\cal O} \rangle^2 ,  \label{hur}
\eea
$\{,\}$ being the anticommutator, and 
\be
\hat{Q} \equiv Q - \langle Q\rangle, ~~~~~~~~~ 
\hat{P} \equiv P - \langle P\rangle. 
\ee

Eq. (\ref{sur}) is the {\it Schr\"odinger Uncertainty Relation} 
\cite{schur}.  Equality  is satisfied by states  $|B,C\rangle$ 
that are colinear in $\hat Q$ and $\hat P$:
\be
\hat{Q} |B,C\rangle = - i~B~ \hat{P} |B,C\rangle, \label{QPco}  
\ee
where $B$ is a complex constant.  
Now going  to wave-function notation, for what will be the 
minimum-uncertainty squeezed states (MUSS), 
Eq. (\ref{QPco}) can be rewritten as \cite{nt2,nagel}
\bea
[Q + iB~ P]\Psi_{ss}(t) &=& C \Psi_{ss}(t), \label{mueq} \\
C &\equiv&  \langle Q \rangle + iB \langle P \rangle . 
\eea 
In general, $B$ and $C$ are both complex.  The solutions 
to Eq. (\ref{mueq}) are  ``squeezed states'' for the system. 

$B$ can be understood to be  the complex squeeze factor by 
i) multiplying Eq. (\ref{QPco}) on the left 
first by $\hat Q$ and then by  $\hat P$ and taking the expectation 
values, ii) then doing the same for the adjoint equation (multiplying on the 
right), and iii) finally using Eq. (\ref{upgen}).  This yields 
\bea 
B &=& i~\frac{ (\Delta Q)^2}{ \langle  \hat{Q} \hat{P} \rangle }
= i~\frac{ \langle  \hat{P} \hat{Q} \rangle }{ (\Delta P)^2},~~~\\
B^* &=& -i~\frac{ (\Delta Q)^2}{ \langle  \hat{P} \hat{Q} \rangle }
= -i~\frac{ \langle  \hat{Q} \hat{P} \rangle }{ (\Delta P)^2},~~~\\
|B|^2 &=& \frac{ (\Delta Q)^2}{ (\Delta P)^2}.
\eea
Thus, $|B|$ yields the relative uncertainties of $Q$ and $P$.

For the particular case $B = 1$, the MUSS are  
the minimum-uncertainty coherent states (MUCS). These  satisfy equality 
of Eq. (\ref{hur}), the {\it Heisenberg Uncertainty Relation}.
That these are coherent states is easily seen from 
Eq. (\ref{mueq}), which then reduces to $\sqrt{2}$ 
times the LOCS Eq. (\ref{locs}).

To intuitively see this, consider Eq. (\ref{mueq}) for the simple
harmonic oscillator.  The solutions are
\be
\psi_{ss}(q) \propto \exp\left[-\frac{1}{2}
\frac{(q - \langle q\rangle)^2}{B} - i \langle p \rangle~q\right].
\ee
$B$ is an extremely complicated functional \cite{n3}
of the complex parameter $\lambda$ of the standard su(1,1) 
squeeze operator:\footnote{
Here, we can allow $\lambda=\lambda(t)$.  
But note that if, in contrast to the present case,
the oprators defining the squeeze operator have time derivatives
(e.g., the su(1,1) operators $M_-$ and $M_+$ of Ref. \cite{newI}),
then this possibility leads to time-derivative 
complications and an inequivalent result.}
\be
S_a(\lambda(t)) = \exp\left[\lfrac{1}{2}\lambda(t)~  a^{\da}a^{\da}
          -   \lfrac{1}{2}\lambda^*(t)~  aa  \right],
       ~~~~~~~~~ \lambda(t) \equiv r(t)e^{i\theta(t)}.
\label{Salambda}
\ee
The exact relationship  depends on if one defines the 
displacement-operator squeezed states (DOSS) as 
\be
|\alpha,\lambda;t\rangle = D_A(\alpha)S_a(\lambda)|0\rangle
\label{ds}
\ee
or as  $S_a(\lambda)D_A(\alpha)|0\rangle$.

We will now describe and then apply this formalism to  
a well-known and important system,
the Paul trap.  The aim is to obtain a deeper insight into the ``coherent''
and ``squeezed'' states of this system.


\section{\textsf{The Paul Trap}} 

The Paul trap is a dynamically stable environment for charged particles
\cite{dawson}-\cite{paulnobel}.  It has been of great use in areas 
from quantum optics to particle physics. 

It's main structure consists of two parts.  The first is an 
annular ring-hyperboloid of 
revolution, whose symmetry is about the $x-y$ plane at  $z=0$.  
The distance from the origin to the ring-focus of the hyperboloid 
is $r_0$.  The inner surface of this ring electrode is a time-dependent  
electrical equipotential surface.  
The second part of the structure 
consists of two end-caps.  These are hyperboloids 
of revolution about the $z$ axis.   The distance from the origin to the 
two foci is usually $d_0 = r_0/\sqrt{2}$.  The two end-cap surfaces are 
time-dependent equipotential surfaces with  sign opposite  to 
that of the ring.  The electric field within this trap is a quadrupole field. 
With oscillatory potentials  are applied, a charged particle {\it can be} 
dynamically stable.  

Paul gives a delightful mechanical analogy \cite{paulnobel}.  Think 
of a mechanical ball put at the center of a saddle surface.  With no 
motion of the surface, it will fall off of the saddle.  However, if the 
saddle surface is rotated  {\it with an appropriate frequency}   
about the axis normal to 
the surface at the inflection point, 
the particle will be stably confined.
The particle is oscillatory about the origin in both the $x$ and $y$ 
directions.  But it's  oscillation in the $z$ direction 
is restricted to be bounded from below by  some $z_0>0$.

The potential energy can be parametrized as \cite{dawson}
\be
V(x,y,z,t) =  V_x(x,t) + V_y(y,t) +V_z(z,t),
\ee
where
\bea
V_x(x,t) &=&  + \frac{e}{2r_0^2}~{\cal V}(t)~x^2
        \equiv \frac{1}{2} \Omega_x(t) x^2, \\
V_y(y,t) &=&  + \frac{e}{2r_0^2}~{\cal V}(t)~y^2
    \equiv \frac{1}{2} \Omega_y(t) y^2, \\
V_z(z,t) &=&  - \frac{e}{r_0^2}~{\cal V}(t)~z^2
    \equiv \frac{1}{2} \Omega_z(t) z^2.  \label{vsubz}
\eea
In the above, 
\be
{\cal V}(t) = {\cal V}_{dc}   -  {\cal V}_{ac}\cos{\omega (t-t_0)}
\ee
is the ``dc'' plus ``ac time-dependent'' electric potential that is 
applied between the ring and the end caps.  
These potentials can be used to solve the classical motion problem.  
The result is  
oscillatory Mathieu functions for the bound case \cite{dawson}.  The 
oscillatory motion goes both positive and negative in the $x-y$ plane, but
is constrained to be positive in the $z$ direction. 

Exact solutions for the quantum case were first investigated in detail by   
Combescure \cite{monique}.  In general, work has concentrated on the $z$
coordinate, but not entirely \cite{chi}.  Elsewhere \cite{ntpaul} we will
look at the symmetries, separations of variables, and the number and 
coherent state solutions of the three-dimensional Paul trap, in both
Cartesian and cylindrical coordinates.  


\section{\textsf{Coherent physics of the Paul trap}} 
\subsection{Coherent states and the classical motion}

With the background established in the previous two sections, we want 
to discuss the coherent/squeezed states
of the Paul trap.  
We focus on the interesting $z$ coordinate and, in
particular, use as reference the lovely discussion in Schrade et al. (SMSG) 
\cite{schrade}.   Using Eq. (\ref{vsubz}), the Hamiltonian is 
\be
H=i\partial_t=-\frac{1}{2}\frac{\partial^2}{\partial z^2}+\frac{1}{2}
\Omega_z(t) z^2.
\ee

The connection between the notation of  Section \ref{notation} 
and that of SMSG \cite{schrade} is 
\be
q=z,~~~~2f=\Omega_z,~~~~\xi(t)=\lfrac{1}{\sqrt{2}}\epsilon(t).
\label{ident}
\ee
The ladder operators, $A(t)$ and $A^{\da}(t)$, are 
\bea
A(t) &=& \frac{i}{\sqrt 2}\left[\ep~p -\epd~z \right]
     =  \frac{1}{2}\left\{a[\ep-i\epd] + a^\da[-\ep -i\epd]\right\},
         \label{A}\\
A^\da(t) &=& -\frac{i}{\sqrt 2}\left[\eps~p -\epsd~z \right]
     =  \frac{1}{2}\left\{a^\da[\eps+i\epsd] + a[-\eps +i\epsd]\right\}, 
         \label{Adag}
\eea
Recall that because of Eq. (\ref{ident}), $\ep$ is a complex solution to 
the differential equation (\ref{ode})
and its complex conjugate, $\eps$, is the other linearly independent solution. 
The Wronskian of $\ep$ and $\eps$ is a constant:
\be
W(\epsilon,\epsilon^*)= \ep \epsd - \epd \eps = -2i. \label{W}
\ee 

Note from Eq. (\ref{vsubz}), that Newton's classical equation of motion is 
\be
F = \ddot{z}_{cl}(t)= -\frac{dV(z,t)}{dz} =    \frac{2e}{r_0^2}~{\cal V}(t)~z
    \equiv -  \Omega_z(t) z_{cl}(t),
\label{newton}
\ee
the solutions being Mathieu functions.  
But Eq. (\ref{ident})  shows that Eq. (\ref{newton}) is exactly the 
form of Eq. (\ref{ode}) for $\ep$ and $\eps$.  
This means that a combination of $\ep$ and $\eps$, up to normalization, 
follows the classical-motion Mathieu
function solutions for $z_{cl}(t)$ and  $p_{cl}(t)=\dot{z}_{cl}(t)$.  
Therefore, we can conveniently and with foresight write these 
classical solutions in the forms most convenient for the quantum study:  
\bea
z_{cl}(t) &=& \frac{i}{2}\left\{\left[\eps\epo-\ep\epso\right]p_o+
\left[\ep\epsdo-\eps\epdo\right]z_o\right\},  \\
p_{cl}(t) &=& \frac{i}{2}\left\{\left[\epsd\epo-\epd\epso\right]p_o+
\left[\epd\epsdo-\epsd\epdo\right]x_o\right\}
           =   \dot{z}_{cl}(t), 
\eea
where $z_o$ and $p_o$ are initial position and momentum  and 
$\epo=\epsilon(t_o)$.  

Now using Eqs. (\ref{QP}), (\ref{A}), and (\ref{Adag}), 
with $Q \rightarrow Z$, 
\bea 
Z(t) &=& \frac{i}{2}\left\{
        [\ep-\eps]p - [\epd-\epsd]z\right   \}     \\
     &=& \frac{i}{\sqrt 2}\left\{
        a~[(Im~ \epd)+i (Im ~\ep)] - a\dg~[(Im~\epd)-i (Im~\ep)]\right\}, \\
P(t) &=& \frac{1}{2}\left\{
        [\ep+\eps]p - [\epd+\epsd]z\right\}   \\
     &=& \frac{1}{i\sqrt2}\left\{
       a~[(Re~\ep)-i (Re~\epd)] - a\dg~[(Re~\ep)+i (Re~\epd)]\right\}.
\eea
It follows that  
\be
[Z(t),P(t)] = -\lfrac{1}{2}\left[\eps\epd-\ep\epsd\right] = i,
\ee
where we have used the Wronskian (\ref{W}).
 
Using Eqs. (\ref{docs}), (\ref{A}), and (\ref{Adag}), one can calculate 
that
the coherent-state wave functions  for the Paul Trap  go as \cite{schrade}
\be
\Psi_{cs}(z,t) \sim \exp\left[-\frac{1}{2}\left(\frac{-i\epd}{\ep}\right)
       \left(z - z_{cl}(t)\right)^2+izp_{cl}(t)\right].  \label{cs}
\ee
The Gaussian form of $\Psi$ can be readily verified 
{\cite{newI,newII-III}} by noting that i)
\bea
\left(-i\frac{\epd}{\ep}\right) &=&\frac{1}{\phi(t)}\left[1-\lfrac{i}{2}
\dot{\phi}(t)\right],\label{RIparts} \\
\phi(t)&=& {\ep}{\eps} \label{phi}
\eea
and ii)  $\phi$ is a positive real function of $t$.\footnote{
The function (\ref{RIparts}) also arises in number-operator 
states.  the relevant equation is $A(t)|0;t\rangle = 0$,   
when $A(t)$ of Eq. (\ref{A}) is used.  
This is the $0$-eigenvalue case of Eq. (\ref{locs}).} 


\subsection{Uncertainty relations}

However, with these wave functions, SMSG \cite{schrade} found that the 
$z-p$ Heisenberg Uncertainty Relation is not satisfied.  Rather, 
the $z-p$ Schr\"odinger Uncertainty Relation is satisfied 
\cite{oldI-II,schrade,gt}:  
\be
(\Delta z)^2(\Delta p)^2 = 
           \lfrac{1}{4}\left[1+\lfrac{1}{4}\dot{\phi}^2(t)\right], 
\label{zpur}
\ee
where $\phi$ is given in Eq. (\ref{phi}).\footnote{
The right hand side of Eq. (\ref{zpur}) is equal to $1/4$ for all 
$t$ only if $\dot{\phi}=0$; that is, for an harmonic oscillator or a 
driven oscillator \cite{gt}.}

We observe that not only is this correct but is to be expected.  These
wave functions were generated by the $Z-P$ variables, and hence should
satisfy the Heisenberg uncertainty relation for $Z$ and $P$.  Contrariwise,
note that Eq. (\ref{A}) can be written as  
\be
A(t) =  \frac{1}{2}\left\{a[\ep-i\epd] + a^\da[-\ep -i\epd]\right\}
     \equiv \mu a + \nu a^\da.     \label{Aoft}
\ee
But with Eq. (\ref{W}) one has that 
\be
    1= |\mu|^2 - |\nu|^2.\label{munu1}
\ee
That is, $A$ and $A^\da$ must be  related to $a$ and $a^\da$ by a 
Holstein-Primakoff/Bogoliubov transformation \cite{hpb} 
of the form
\bea
S_a^{-1}(\lambda)~a~S_a(\lambda) &=& [\cosh r]a + [e^{i\theta}\sinh r]a^\da,
\label{sas}\\
S_a^{-1}(\lambda)~z~S_a(\lambda) &=& \frac{1}{\sqrt 2}\left\{
\left[\cosh r +e^{-i\theta}\sinh r\right]a + 
\left[\cosh r +e^{i\theta}\sinh r\right]a^{\dagger} \right\}. \label{szs}
\eea

However, there remains one further complication.  The coefficient of $a$ 
on the right of  Eq.  (\ref{sas}) is real while that of Eq. (\ref{Aoft}) is 
complex.  There is a phase offset.  This is related to  the fact that there
are four parameters in 
Eq. (\ref{Aoft}) and only two in Eq.  (\ref{sas}).  That there are only 
two parameters in Eq (\ref{sas}) follows 
from the fact that the entire squeezed-state transformation 
from $(z,p)$ to $(Z,P)$ also involves 
a displacement. (See Eq. (\ref{ds}).)
The displacement supplies the other two parameters, as we now demonstrate.  


\subsection{Squeezed states}

Combine, in the $A$ representation, the DOCS definition of Eq. (\ref{docs}) 
with the LOCS definition of Eq. (\ref{locs}): 
\be
A(t)D_A(\alpha)|0;t\rangle = \alpha D_A(\alpha)|0;t\rangle.  
\label{csA}
\ee
Next, insert $I= S_a^{-1}(\lambda)~S_a(\lambda)$ in front of all the operators
and multiply on the left by $S_a(\lambda)$  of Eq. (\ref{Salambda}).
Regrouping,  we obtain the equation
\be
\left(S_aAS_a^{-1}\right)
\left(S_aD_A(\alpha)S_a^{-1}\right)S_a(\lambda)~|0;t\rangle
=\alpha\left(S_aD_A(\alpha)S_a^{-1}\right)S_a(\lambda)~
|0;t\rangle. \label{t2}
\ee

Given Eqs. (\ref{Aoft}) and (\ref{munu1}) 
and using both the BCH relation {\cite{oldI-II}} 
\be
\exp\left[\lambda\frac{a^\da a^\da}{2}-\lambda^*\frac{aa}{2}\right] = 
\exp\left[\gamma_+ \frac{a^\da a^\da}{2}\right]
\exp\left[\gamma_3 (a^\da a + \lfrac{1}{2}) \right]
\exp\left[\gamma_- \frac{aa}{2},         \right],\label{suBCH1} 
\ee
\be
\gamma_-=-e^{-i\theta}\tanh{r},~~~~\gamma_+=
e^{i\theta}\tanh{r},~~~~\gamma_3=\ln{\cosh{r}},
\label{suBCH2}
\ee
and also the theorem {\cite{Miller}}
\be
\exp[X]\exp[Y]\exp[-X]=\exp[e^{X}Ye^{-X}],\label{thm}
\ee
it follows that
\be
S_a(\lambda)D_{A}(\alpha)S_a^{-1}(\lambda) = 
\exp\left[\beta a^{\da} - \beta^* a\right] \equiv D_{a}(\beta), 
\label{doa} 
\ee
\be
\beta = \alpha v^*-\alpha^* u,~~~~~\beta^* = \alpha^*v-\alpha u^*.
\label{abl}
\ee
In addition, we see that 
\be
S_a(\lambda)AS_a^{-1}(\lambda) = va+ua^{\da}, \label{SASda}
\ee
where the coefficients $u$ and $v$ are
\be
u =  \nu\cosh{r}-\mu e^{i\theta}\sinh{r},~~~~
v  =  \mu\cosh{r}-\nu e^{-i\theta}\sinh{r}.
\label{uv}
\ee
Note that $u$ and $v$ are functions of $t$ because $\mu$ and $\nu$ are.
Furthermore, we  have
\be
|v|^2-|u|^2 = |\mu|^2-|\nu|^2 = 1.
\ee

Therefore, Eq. (\ref{t2}) becomes
\footnote{
In Eq. (\ref{T3}), $\alpha$ is indeed the quantity of Eq. (\ref{csA}).   
However, because of our construction, it can be obtained from Eqs. 
(\ref{abl}) and (\ref{uv}) as a functional of  $\lambda$ and $\beta$.}
\be
\left\{va+ua^{\da}\right\}~ 
D_{a}(\beta)~S_a(\lambda)~|0;t\rangle 
= \{ \alpha(\lambda,\beta)\} ~ D_{a}(\beta)~ S_a(\lambda)~|0;t\rangle
= \{\alpha(\lambda,\beta)\} ~ |\alpha(\lambda,\beta),t\rangle.     
   \label{T3}
\ee
This satisfies both the displacement-operator definition of squeezed states
(to the right of the  brackets) and the ladder operator definition of  
squeezed states (the curly brackets) \cite{newI}.   
This completes the 
transformation of a coherent state in the $(A,A^{\da})$ representation into 
a squeezed state in the $(a,a^{\da})$ representation.  
Since the squeeze 
operator is invertible, we can obviously reverse this process.
\footnote{
Because of the isomorphism of the two Heisenberg-Weyl algebras, 
$\{a,a^{\da},I\}$ and $\{A(t),a^{\da}(t),I\}$, the converse of this 
result follows.  Starting with a coherent-state equation  
in the $(a,a^{\da})$ representation: 
$
aD_{a}(\eta)|0\rangle =\eta D_{a}(\eta)|0\rangle,
$
where $\eta$ is complex,     
we can map this equation into an analogous squeezed-state equation in 
the $(A,A^{\da})$ representation 
with an appropriate squeezing operator in the $(A,A^{\da})$ representation. 
This is done going through an analogous procedure as above with the aid of 
the equations
$
a=\rho A + \sigma A^{\da}, 
$
$
\rho = \lfrac{1}{2}\left(\eps - i\epsd\right),
$
and
$
\sigma = \lfrac{1}{2}\left(\ep - i\epd\right)
$
and the condition
$
|\sigma|^2-|\rho|^2=1.
$
}


\section{\textsf{Conclusion}}

The coherent states of  
$(Z,P)$ or $(A,A^\da)$ are squeezed states of $(z,p)$ or $(a,a^\da)$, 
and {\it vice versa}.
This makes sense.  The fundamental potentials are of different widths, 
so their
coherent-state  Gaussians are also of different widths.  Indeed,  
Eq. (\ref{cs}) shows this.  For the $(z,p)$ uncertainty relation, 
$[{-i\epd}/{\ep}]$ is a squeeze factor.  

~

\section*{\textsf{Acknowledgements}}

MMN acknowledges the support of the United States Department of 
Energy and the Alexander von Humboldt Foundation.  DRT acknowledges
a grant from the Natural Sciences and Engineering Research Council 
of Canada.



\end{document}